\documentclass[12pt]{article}
\usepackage{amsmath}
\usepackage{amsfonts}
\usepackage{amssymb}
\usepackage{latexsym}
\usepackage{graphicx}
\begin{document}
\input epsf

\begin{flushright}
OHSTPY-HEP-T-04-001\\
hep-th/0401115
\end{flushright}
\vspace{20mm}
\begin{center}
{\LARGE  Where are the states of a black hole?\footnote{Talk given at `Quantum Theory and Symmetries',
Cincinnati, September 2003.}}
\\
\vspace{18mm}
{\bf   Samir D. Mathur
\\}
\vspace{8mm}
Department of Physics,\\ The Ohio State University,\\ Columbus,
OH 43210, USA\\ 
\vspace{1mm}
mathur@mps.ohio-state.edu
\vspace{4mm}
\end{center}
\vspace{10mm}
\thispagestyle{empty}
\begin{abstract}

We argue that bound states of branes have a size that is of the same order as the horizon radius of the corresponding black hole. Thus the interior of a black hole is not `empty space with a central singularity', and Hawking radiation can pick up information from the degrees of freedom of the hole.

\end{abstract}
\newpage

\section{Introduction}

Let us recall the puzzles arising in the quantum physics of black holes. Gedanken experiments show that
we must associate an entropy \cite{bek}
\begin{equation}
S={A\over 4G}
\label{one}
\end{equation}
with a black hole, where $A$ is the area of the horizon. Statistical mechanics then suggests that there should
be $e^S$ states of the hole. But the geometry of the hole appears to be completely determined by 
its mass, charge and angular momentum, implying an entropy $S=\ln 1=0$, which contradicts (\ref{one}).
Closely related is the `information paradox': Hawking radiation from the hole is created by the progressive dilation
of vacuum modes near the horizon, and if the geometry here is determined only by the above conserved quantities then
the radiation will carry no details of the matter which went in to make the hole. We thus get a violation of the unitarity
of quantum mechanics \cite{hawking}.

Clearly, the key to resolving these long standing problems is to understand where the
states of the hole are located; in other words, to understand the `hair' that differentiate
different states of the hole. In this talk we propose that the interior of the horizon is not described 
by the conventional picture of Fig.1(a) where we have `empty space with a central singularity'. Rather
the differences between the $e^S$ states are manifested throughout the interior of the `horizon' (Fig.1(b)),
there is in general no special point to play the role of the singularity,
and the horizon is just the boundary of the region where the typical states differ from each other.

\begin{figure}[ht]
\centerline{\epsfxsize=3.9in\epsfbox{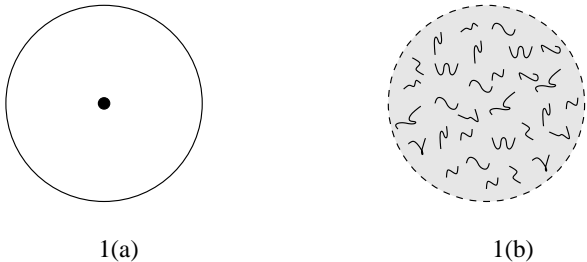}}   
\caption{(a) The conventional picture of a black hole \quad (b) the proposed picture -- state information is distributed throughout the `fuzzball'. \label{inter}}
\end{figure}

We will first give an abstract argument for the latter picture. Next, we show that the simplest string theoretic
system with entropy -- the 2-charge extremal D1-D5 system -- behaves like Fig.1(b) rather than Fig.1(a); here we can construct all the $e^S$ states explicitly in the gravity description. We then construct one state of the
3-charge extremal system -- a state with large D1, D5 charges and just one unit of momentum P.
This state also has the structure of Fig.1(b).  Lastly, we describe the physics which might underlie this large departure from the 
conventional picture Fig.1(a). String states exhibit entropy when different types of branes bind and the degrees of freedom are carried by `fractional brane excitations'. But such fractional branes have very low tension and can stretch upto distances of
the order of the horizon radius, destroying the naive picture of the black hole interior.

\section{An abstract argument about `hair'}

A large set of Gedanken experiments have argued for a general link between entropy and horizons --
whenever we find a horizon with area $A$ then we must associate an entropy $S={A\over 4G}$ with the region behind
 that horizon. But now consider the $e^S$ microstates that give the entropy of a black hole. Some attempts to locate the
 `hair' have looked for small perturbations near the horizon of Fig.1(a).  Other approaches would suggest that the microstates differ only within a planck distance of the central singularity. In either case each microstate looks pretty much like
 Fig.1(a), in that it has a horizon with an area $\approx A$. 
 
 But if each of the $e^S$ microstates has such a horizon, then it must itself represent $\sim e^S$ different states. This makes no sense, since we wanted the microstates to explain the  entropy, not have further entropy themselves. We conclude that
 {\it the $e^S$ microstates should have no horizons individually -- the notion of a horizon should arise only after `coarse-graining' over these microstates}.
 
 \section{The D1-D5 system}
 
 This looks like a tall order since to avoid having a horizon the microstates must differ from the naive black hole geometry not just within planck distance of the singularity but all the way upto the horizon. Remarkably, however, for the  extremal D1-D5 system we found exactly this  structure. We take type IIB string theory compactified on $T^4\times S^1$. The volume of $T^4$ is $(2\pi)^4 V$ and the length of $S^1$ is $2\pi R$. We wrap $n_5$ D5 branes on $T^4\times S^1$ and $n_1$ D1 branes on $S^1$. The metric usually written for this system (which we will call the `naive metric') is
 \begin{equation}
ds^2={[-dt^2+dy^2]\over \sqrt{(1+{Q_1\over r^2})(1+{Q_5\over
r^2})}}+\sqrt{(1+{Q_1\over r^2})(1+{Q_5\over
r^2})}dx_idx_i+\sqrt{{1+{Q_1\over r^2}\over 1+{Q_5\over r^2}}}dz_adz_a
\label{naive}
\end{equation}
and is sketched in Fig.2(a). 
One knows that this geometry cannot be the complete story, since the branes wrapped as above give a 1+1 dimensional CFT with fermions in the Ramond sector, and the ground state here has degeneracy $e^{2\sqrt{2}\pi\sqrt{n_1n_5}}$. We must therefore see a corresponding degeneracy in the dual gravity description. By a sequence of S,T dualities we map the D1-D5 system to the FP system -- the fundamental string (F) wrapped $n_5$ times on $S^1$, carrying $n_1$ units of momentum (P) along $S^1$.  Since we are counting the {\it bound} states of the FP system the P charge is carried on the F string as traveling waves. But the F string has no longitudinal vibrations, so it must bend away from its central axis in a {\it transverse} direction to carry the momentum. The $n_5$ strands of the F string separate under this deformation and spread over a nonzero transverse region, thus cutting off the naive FP metric before $r=0$. Dualizing back we get (classically) a family of geomeries for the D1-D5 system characterized by $\vec F(v)$, the vibration profile of the F string \cite{lm4}
\begin{equation}
ds^2={-(dt\!-\!A_i dx^i)^2\!+\!(dy\!+\!B_i dx^i)^2\over \sqrt{[(1+K)/H]}}+\sqrt{1+K\over
H}dx_idx_i+\sqrt{H(1+K)}dz_adz_a
\label{six}
\end{equation}
\begin{eqnarray}
e^{2\Phi}&=&H(1+K), ~~~C^{(2)}_{ti}={B_i\over 1+K},
~~~C^{(2)}_{ty}=-{K\over 1+K}\nonumber\\
C^{(2)}_{iy}&=&-{A_i\over 1+K},
~~~~C^{(2)}_{ij}=C_{ij}+{A_iB_j-A_jB_i\over 1+K}
\label{twenty}
\end{eqnarray}
where $B_i, C_{ij}$ are given by
\begin{equation}
dB=-*_4dA, ~~~dC=-*_4dH^{-1}
\end{equation}
and $*_4$ is the duality operation in the 4-d transverse  space
$x_1\dots
x_4$ using the flat metric $dx_idx_i$. The functions $H,K, A_i$ are generated by $\vec F$
 \begin{equation}
{1\over H}=1+{Q\over L}\int_0^L\!\!\!\!\! {dv\over |\vec x-\vec F(v)|^2}, ~K={Q\over
L}\int_0^L \!\!\! {dv (\dot
F(v))^2\over |\vec x-\vec F(v)|^2},
~A_i=-{Q\over L}\int_0^L\!\!\!\!\! {dv\dot F_i(v)\over |\vec x-\vec F(v)|^2}\\
\label{functions}
\end{equation}
The geometries, pictured in Fig.2(b), have no horizons and no singularity (at points $\vec x=\vec F(v)$
we have just the coordinate singularity arising at the center of Kaluza-Klein monopoles \cite{lmm}). If we `coarse-grain'
by drawing a boundary to enclose the region where these geometries differ significantly from each other
(the dashed line in Fig.2(b)) then from the area $A$ of this boundary we find \cite{lm5}
\begin{equation}
S\equiv{A\over 4G}\sim\sqrt{n_1n_5}\sim S_{micro}
\end{equation}
Thus the Bekenstein `area entropy' arises directly as a `coarse-graining over hair'. It is important that the geometries
(\ref{six}) can be distinguished from each other by dynamical experiments. The CFT state dual to the geometry 
described by $\vec F(v)$ is given by the map
\begin{equation}
(a^{i_1}_{n_1})^\dagger\dots (a^{i_k}_{n_k})^\dagger |0\rangle~\rightarrow~\sigma_{n_1}^{i_1}\dots \sigma_{n_k}^{i_k}|0\rangle_{NS}
\end{equation}
where the vibration profile $\vec F(v)$ of the F string is described at the quantum level by transverse oscillators $(a^{i}_{n})^\dagger$
and $\sigma_n^i$ are twist operators of the CFT which is a sigma model with target space the orbifold $(T^4)^N/S^N$. A quantum thrown into the `throat' region of one of these geometries bounces off the end and returns in a time $\Delta t_{sugra}$ which depends on the geometry. The CFT dual of this process has a pair of vibrations traveling around the `multiwound' circle produced by the twist operators, and this takes a time $\Delta t_{CFT}$. We find $\Delta t_{sugra}=\Delta t_{CFT}$, and observe that in this process the backreaction of the test quantum of the geometry remains small \cite{lm4}. Thus we can reliably probe the structure of the `hair'.

\begin{figure}[ht]
\centerline{\epsfxsize=3.9in\epsfbox{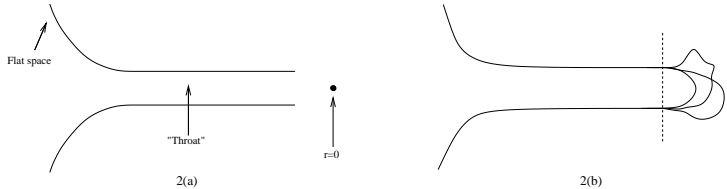}}   
\caption{(a) The naive geometry of extremal D1-D5 \quad (b) the actual geometries; the dashed line gives a `horizon' whose area gives the entropy. \label{inte}}
\end{figure}

\section{The three charge D1-D5-P system}

The `naive' geometry of the 3-charge system also has an infinite length throat extending to a horizon at $r=0$, and this geometry can be extended to a region `inside' the horizon. But  the 2-charge results suggest that the throat gets capped off before $r=0$, with different caps giving different microstates as in Fig.2(b). While we were unable to construct the generic 3-charge state  we note that if our picture were true then we should be able to add a small perturbation to any state of the 2-charge extremal system such that we get  a 3-charge extremal system with one unit of P. It is nontrivial that such a perturbation exist because we demand smoothness everywhere and decay at infinity. Such a perturbation was indeed constructed \cite{mss} (by matching interior and exterior solutions to several orders) giving the geometry dual to the CFT state
\begin{equation}
|\psi'\rangle=J^-_{-1}|\psi\rangle_ R, ~~~(L_0-\bar L_0)|\psi'\rangle=1
\end{equation}
where $|\psi\rangle_R$ is a particular Ramond ground state of the 2-charge system. We thus get the analogue of Fig.2(b) rather than Fig.2(a) for at least this 3-charge state. If the throat is similarly `capped' for generic 3-charge states then we cannot just fall through the horizon at $r=0$ into the black hole interior in the manner implied by the naive geometry analogous to Fig.1(a).

\section{Fractionation}

Consider a string wrapped on a circle of length $2\pi R$. The minimum  excitation (with no net P charge) comes from one left and one right mover, with $\Delta E={1\over R}+{1\over R}={2\over R}$. But if we have a {\it bound } state of $n$ strings, then we get effectively one string of length $2\pi nR$. The momentum modes then come in fractional units ${1\over nR}$ and the threshold drops to \cite{dm}
$\Delta E={2\over nR}$. Similarly, the threshold for vibrating a D1-D5 bound state is \cite{ms} ${2\over n_1n_5 R}$.
 We see that bound states made out of  many branes have very light `fractional' excitations.
 
 To see the physical consequence of such excitations, consider an extremal 3-charge D1-D5-P bound state and imagine  a test quantum brought  to a distance $r$ from this state. This needs a minimum localization energy $\sim 1/r$, which is small if $r$ is macroscopic. But the fractional excitations of the bound state are light too, and we ask if using the energy $1/r$ a pair of  fractional brane excitations can be created that extend from the origin to the location $r$ of the test quantum. We want this process to be probable and not just possible, so we demand that the entropy gained by creating the fractional pair be at least $\Delta S=1$ (thus the phase space increases by a factor $e\approx 2.7$ upon creation of the pair). A standard entropy computation then yields that $\Delta S>1$ for $r<r_{crit}$ where \cite{emission}
 \begin{equation}
 r_{crit}~\sim ~[{g^2\alpha'^4\sqrt{n_1n_5n_p}\over VR}]^{1\over 3}~\sim ~r_H
 \label{twentyqq}
 \end{equation}
 where $r_H$ is the horizon radius of the D1-D5-P bound state! This estimate supports the `fuzzball' picture Fig.1(b), and is
 in line with the above construction of `hair'  for D1-D5 and D1-D5+(1 unit of P).
 
 As with all string theory constructions of black holes, we expect that if extremal holes  behave like
 Fig.1(b) then near extremal and generic black holes will be qualitatively similar. The entropy for general black holes in 5-D is reproduced {\it exactly} if we partition the mass optimally among brane-antibrane pairs \cite{hms}
 \begin{equation}
 S=2\pi(\sqrt{n_1}+\sqrt{\bar n_1})(\sqrt{n_5}+\sqrt{\bar n_5})
(\sqrt{n_p}+\sqrt{\bar n_p})
\end{equation}
Assuming such a  microstructure the size estimate along the lines of (\ref{twentyqq}) again gives $r_{crit}\sim r_H$.

\section{Conclusion}

Hawking's calculation of radiation showing information loss is so robust because it uses no details of the physics at
the planck scale. Resolving the information paradox thus needs an explicit nonlocality over macroscopic distances.
String theory has succeeded in matching  entropy and radiation rates of brane bound states  with corresponding black holes \cite{stromvafa}. Using these same systems we have argued that bound states swell upto a size of order the horizon radius; thus the interior of the horizon is
not just `empty space with a central singularity'. This makes it possible for  radiation from the hole to pick up information from the `hair' and avoid the information paradox.

It will be interesting to address dynamical questions with the above picture.  It is possible that infalling matter falls straight through the `fuzzball' towards $r=0$ (as if it were falling through a conventional horizon), but over  the Hawking evaporation time information is transferred to the `light fractional modes' and into the radiation.


\section*{Acknowledgments}

Much of the work described here is in collaboration with O. Lunin, A. Saxena and Y. Srivastava.
I  thank for discussions I. Bena, S. Das, A. Dabholkar, V. Hubeny, C. Johnson, P. Kraus, D. Kutasov,  E. Martinec,
S. Shenker and L. Susskind. I thank A. Saxena for making the figures. 
This work was supported in part by DOE grant DE-FG02-91ER-40690.



\end{document}